\documentclass[aps,pre,superscriptaddress,twocolumn,floatfix]{revtex4}

\usepackage{graphicx}

\begin{document}

\title{Magnetically controlled ballistic deposition. \\A model of polydisperse
granular packing}

\author{K. Trojan},
\affiliation{SUPRAS and GRASP\footnote{GRASP = Group for Research in Applied
Statistical Physics;\\ SUPRAS = Services Universitaires Pour la
Recherche et les Applications en Supraconductivit\'e}, Institute of Physics, 
B5, University of Li$\grave e$ge,\\ B-4000 Li$\grave e$ge, Euroland}
\affiliation{Institute of Theoretical Physics, University of Wroc\l{}aw,\\ 
pl. M. Borna 9, 50-204 Wroc\l{}aw, Poland}
\author{M. Ausloos}
\affiliation{SUPRAS and GRASP\footnote{GRASP = Group for Research in Applied
Statistical Physics;\\ SUPRAS = Services Universitaires Pour la
Recherche et les Applications en Supraconductivit\'e}, Institute of Physics, 
B5, University of Li$\grave e$ge,\\ B-4000 Li$\grave e$ge, Euroland}

\begin{abstract}
The flow and deposition of polydisperse granular materials
is simulated through the Magnetic Diffusion Limited Aggregation (MDLA)
model. The random walk undergone by an entity in the MDLA model is
modified such that the trajectories are ballistic in nature, leading to
a magnetically controlled ballistic deposition (MBD) model. This allows
to obtain important ingredients about a difficult problem that of the
nonequilibrium segregation of  polydisperse sandpiles and heterogeneous
adsorption of a binary distribution of particles which can interact with
each other and with an external field. Our detailed results from many
simulations of MBD clusters on a two dimensional triangular lattice
above a flat surface in a vertical finite size box for binary systems
indicates intriguing variations of the density, ''magnetization'', types
of clusters, and fractal dimensions. We derive the field and grain
interaction dependent susceptibility and compressibility. We deduce a
completely new phase diagram for binary granular piles and discuss its
complexity inherent to different grain competition and cluster growth
probabilities.
\end{abstract}

\maketitle

\section{ INTRODUCTION}

Understanding the flow and static structures of granular matter is
becoming increasingly relevant. Many raised questions are tackled along
various lines of approach\cite{granularref}. It has already been claimed
that the  simplifications found in basic models need to be improved in
order to explain features of such complex materials. Granular pile
spreading processes driven by cooperative non-linear evolution rules
lead to developed patterns which often reach a high level of complexity.
It is of present interest to examine whether growth models can be used
for describing granular structures and related material properties.
Furthermore the non-linear processes at work in granular flows and
depositions hint toward simulation approaches\cite{Pandey}. Cooperative
effects  in ballistic deposition of hard disks have been recently
mentioned\cite{Faraudo}.

Kinetic growth models (KGM) have received much
attention,\cite{Herrmann} like the Eden model\cite{eden} and the
diffusion limited aggregation model\cite{DLA} (DLA). They have served to
describe nonequilibrium phenomena like  film or crystal
growth\cite{crystalgrowth1,crystalgrowth2}, epidemics\cite{epidemics},
material fractures\cite{materialfractures}, ... In all cases, such
models are mainly concerned by the transition from dense branching to
dendritic morphology.

One important physical constraint has to be  considered in describing
granular materials: the materials are not made of symmetrical (spherical
or cubic) entities. The surface of grains is usually rough, thus leading
to  specific angles of repose\cite{angleofrepose}. Also the grain
anisotropy leads to  phenomena like jams\cite{jams}, in flow, and
arches\cite{arches}, in static  structures. It seems therefore necessary
to have {\it at least} one degree of freedom in order to describe
grains; we are even aware that $only$ $one$ $degree$ is a very strong
approximation. This degree of freedom should be coupled to some  field,
just like a spin to a magnetic field. Whence one can imagine that grains
are identical entities except for one degree of freedom, call it a
$spin$ though it  can be any physical feature of particular interest,
like the grain roughness or  shape feature. Clearly a spin allows for
referring to a direction or a rotation process; if this is admitted, to
take such a degree of freedom into account in describing granular piles
should basically improve the granular state overall description.
(Generalizations are immediately imagined by anyone familiar with spin
models and statistical mechanics; one can later on imagine many
component vector models, including Potts-like models\cite{Potts}). In
fact a constrained Ising spin chain has been recently considered and
studied as a {\it toy model} for granular compaction\cite{slow}. The
exchange energy $J$ describing the ''spin-spin interaction'' is
analogous in granular matter to the contact energy due to surface
roughness between grains. A similar interpretation of $J$ for $flows$
can be found in Pandey et al.\cite{Pandey}. The {\it external
''magnetic'' field} in such a case can be e.g. a wind field, the sign
depending on e.g. change in  pressure due to grain drag. We thus combine
topology and mass (or weight) in order to describe granular materials in
a simple way.

The above ideas remind us that a similar set of considerations has  been
found in the magnetic Eden model\cite{MEM} and in the Magnetic
Diffusion Limited Aggregation (MDLA) model\cite{MDLA} when attempting to
describe  crystal growth in a magnetic field, when there is a
competition between  entities.  Aggregation can
proceed under short range or long range dipolar  interactions in
fact\cite{MDipolDLA1,MDipolDLA2}. However the  studies pertained  to the
growth of clusters starting from a point seed. In a recent set of
investigations on the magnetic Eden model, Albano et
al.\cite{Albano0,Albano1,Albano2,Albano3} have pointed out the interest of such
models for examining deposition and film growth, thus starting from a
substrate, - sometimes with rather complex realistic rules. The studies
are also related to nonequilibrium wetting  questions,\cite{Candia} and
other deposition problems.\cite{Cerdeira}

The same type of studies can  be done with the MDLA, i.e. examining  the
growth of clusters from a substrate. It is clear that a substrate having
finite (or not) size destroys the spatial isotropy or more generally
spatial  symmetry. In simple words, there is a top and bottom, if the
substrate is horizontal. It  is obviously of  interest for granular
materials in a gravitational field to consider what happens only  in the
half space $above$ the (finite size of course) substrate. This reduction
is however of fundamental and practical interest because one can also
consider that the system is in a {\it vertical box with walls}, as in
the Albano et al. film growth geometry\cite{Albano1,Albano2,Albano3}
with binary competing entities. In DLA, the diffusing particle follows a
random walk.\cite{DLA} However, it seems very hard to let this usual DLA
rule holds here concerning the path of the granular entity launched far
away from the seed or substrate. In the present considerations, it seems
more appropriate to let the granular entity follows a ballistic vertical
trajectory like in rain models\cite{Cerdeira} rather than a random walk.

In this paper, we calculate what changes result in the features of a
classical ballistic deposition\cite{Cerdeira,BDM,Meakin} (BD) model when we add one
extra degree of freedom, a ''spin'', to the classical BD. The spin
reflects e.g. the $orientation$ of oblate grains or their $roughness$
characteristic.  The spins, as usual, interact through some exchange
energy $J$ which, for grains, is often mechanical or electrostatic in
nature. The external field $H$ is thought to be the image of a classical
field positioning or influencing the flow of grains, like some wind
velocity or, more generally, pressure difference, or an electric field.
This model is hereby called the magnetically controlled ballistic
deposition (MBD) model falls into the category of kinetic growth models.
In Sect.2, we enumerate the algorithm rules and briefly comment upon
them. In Sect. 3 we present results on the ''density'' and
''magnetization'',  ''susceptibility'' and ''compressibility''. The
types of clusters and their fractal dimensions are discussed in Sect. 4.
In Sect. 5, a brief conclusion can be found.

\section{ EXPERIMENTAL PROCEDURE}

\begin{figure}
\begin{center}
\includegraphics[height=3cm]{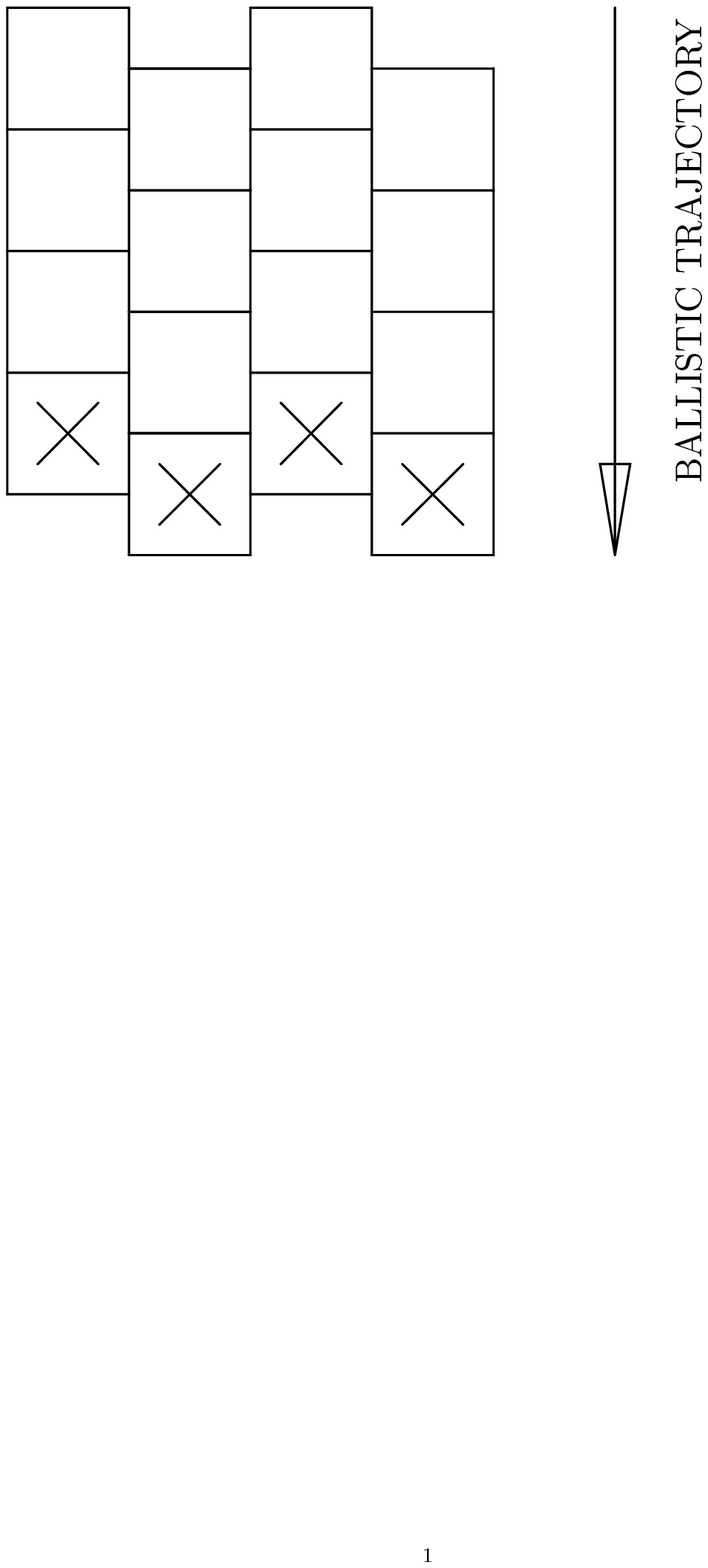}
\end{center}
\caption{\label{figlattice} Example of a 4x4 triangular lattice. The arrow
indicates the fall
direction. The substrate has been marked with crosses.}
\end{figure}

For simplicity we will thereafter call a grain, a spin. It will take
here only two states (up or down) or two values (+1 or -1). The external
field  is supposed to be constant and uniform throughout the whole
system. For obvious  reasons, like higher packing considerations and
possible geometrical frustrations, the underlying lattice should not
have a square symmetry; we have taken a  triangular lattice in the
following simulations (Fig.\ref{figlattice})

One of the main problems to be tackled  is how to choose the best rule
for aggregating granular falling entities (spins) in order to get them
stick together and form  clusters. We have chosen the simplest sticking
rule, usual in statistical physics, i.e., namely a Metropolis-like rule,
as also in Pandey et al.\cite{Pandey}, presented here below. The
algorithm goes as follows:

\begin{enumerate}

\item first, we choose a horizontal substrate of spins with a
predetermined (for example antiferromagnetic-like or random) configuration;

\item{\label{step2}} a falling (up or down) spin is dropped along one of the
lattice lines (see Fig.\ref{figlattice}) from a height $r_{max}+5a$,
where $r_{max}$ is the largest distance between a cluster site and the
substrate i.e. here it is the height of the highest column growing from
the substrate on  the lattice; at each step down the spin can flip i.e.,
change its $sign$,  with equal probability; e.g. the anisotropic grain
can rotate;

\item the spin goes down until it reaches a site perimeter of the cluster; the
local gain in the Ising energy

\begin{equation}\label{isingham}
\beta E = -\beta J \sum_{<i,j>} \sigma_i
\sigma_j - \beta H \sum_{i} \sigma_i,
\end{equation}

is calculated before and after the spin possible impact, thus cluster
growth. If the gain is negative the spin sticks to the cluster
immediately (sticking probability =1.0) and we go back to step
(\ref{step2}). In the  opposite case the spin sticks to the cluster with
a rate $\exp (-\Delta \beta E)$ where $\Delta \beta E$ is the
local gain in the Ising energy. If the spin does not stick to the
cluster it continues going down toward the substrate or bottom of the
box. Of course if the site just below the spin is occupied the spin
immediately stops and sticks to the cluster. When the spin sticks to
the cluster we go back to step (\ref{step2}).

\item After dropping a (large) number of spins the physical quantities of
interest  like the magnetization, density, fractal dimension,  etc.  are
computed. 
\end{enumerate}

It should be noticed that there is no toppling nor relaxation at this
time like in Manna sandpile model or its extensions\cite{slow,Manna,Dickman}.
Moreover, since the number of nearest neighbors on a triangular lattice
is equal to  $6$, and due to the rule of MBD (above paragraph), we do
not have to take into  considerations the spin configurations in which
the depositing spin has a neighbor {\it just vertically over} it or just
below  (because it would then always stick to the cluster). Therefore
there are  only 4 neighboring sites where spin  configurations are
relevant for calculating the local gain in the Ising energy. On the
other hand there are 3 kinds of  site occupation : spin up, spin down
and no spin, hence   52 configurations (excluding the empty one -- when
there is no spin on the perimeter). Some of these configurations are
symmetrical with respect to rotations. Finally 23 configurations are to
be examined having at  least one spin on the perimeter. All these
sticking configuration rates (also distinguishing the sign of
the falling spin)  are shown in table \ref{tableconf}.
The contribution to the sticking rate arising from the
interaction of the depositing spin with the field has also to be
evaluated.

\begin{table*}
\begin{center}
\includegraphics[height=9.5cm]{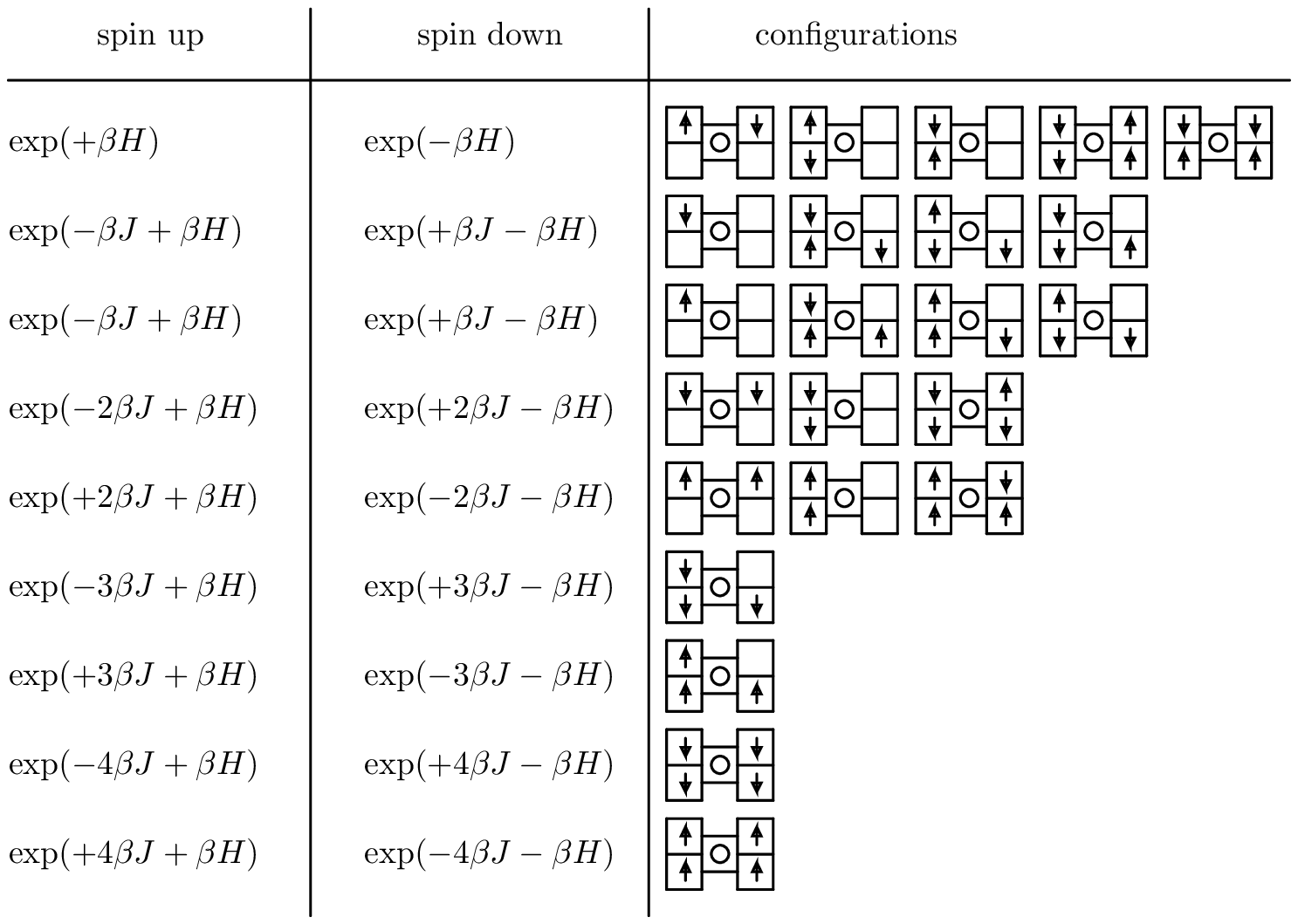}
\end{center}
\caption{\label{tableconf} Rate of sticking and configurations
in the MDM model on the triangular lattice. First (second) column shows
the rate of sticking an up (down) spin to the preexisting  spin
configuration.}
\end{table*}

\begin{figure}
\begin{center}
\includegraphics[height=7cm, angle=-90]{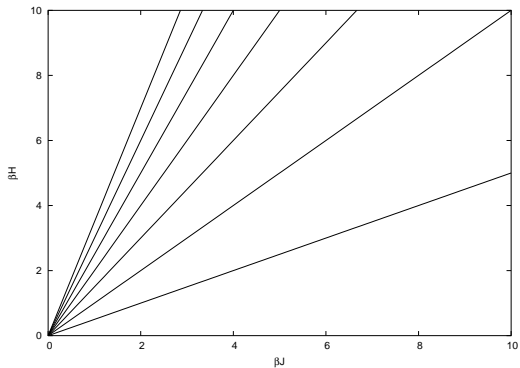}
\end{center}
\caption{\label{figrelations} Relations between $\beta J$ and $\beta H$ on the
$\beta J$-$\beta H$ plane;  $32$ regions result from these relations.}
\end{figure}

The rates of sticking to the cluster read like:

\begin{equation}\label{prob}
P_{n,s} = e^{ - \Delta \beta E} = e^{ s ( n \beta J+ \beta H)},
\end{equation}

where $\Delta \beta E$ is the local gain of the Ising energy, and $n$ is
the difference between the number of up  and down spins: the possible
values are $-4,-3,..,4$, and $s$ is the value (or sign) of the falling
spin($-1$ or $1$). Equating all these rates lead us to a set of
$16$ relations between $\beta J$ and $\beta H$; in fact as in the
MDLA\cite{MDLA}. Fig.\ref{figrelations} determine $32$ regions where
granular packing cluster growth processes differ from each other. The
case $\beta J = 0$ and  $\beta H = 0$ corresponds to the usual ballistic
deposition model (standard  deposition without spin and
field).\cite{Herrmann}

Notice that  the model contains two order parameters, as in the
Blume-Emery-Griffiths model\cite{BEG,ACKP}, one corresponds to the
density, the other to the magnetization.

\section{ NUMERICAL RESULTS and DISCUSSION}

All results reported below are for a triangular lattice of  horizontal size
$L=100$, i.e. the width of the seed substrate,  and when the pile made of
clusters has reached a $500$ lattice unit height.  Every reported data point
corresponds to an average over $1000$ simulations.

\subsection{ Density}

\begin{figure}
\begin{center}
\includegraphics[height=8cm, angle=-90]{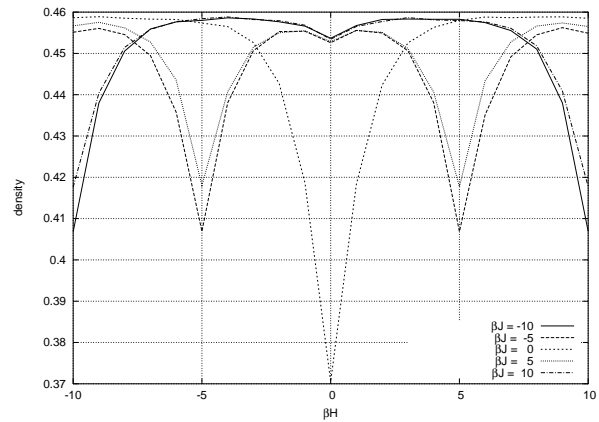}
\end{center}
\caption{\label{figdj} Dependence of density on $\beta H$ for different $\beta
J$. Observe that the behavior is symmetrical with respect to $\beta H = 0$.}
\end{figure}

We define the density of a cluster as
\begin{equation}
G = \frac{\mbox{number of spins in the cluster}}{\mbox{number of sites on the
lattice}},
\end{equation}
in which obviously the number of lattice sites $= 50 000$.
Fig.\ref{figdj} illustrates the behavior of the density with respect to
the $\beta H$  parameter. This figure convinces us that the results are
symmetrical with  respect to $\beta H = 0$. Therefore in the following
subsections we will often present  results for $\beta H > 0$ only. It is
observed that the density presents a sharp  minimum when $\beta H = \pm
\beta J$. The granular pile is rather loosely packed since the density
varies between $0.37$ and $0.45$. This low value with respect to
experimental findings arises from the fact that we have not included
relaxation processes in this investigation.

\begin{figure}[ht]
\begin{center}
\includegraphics[height=8cm, angle=-90]{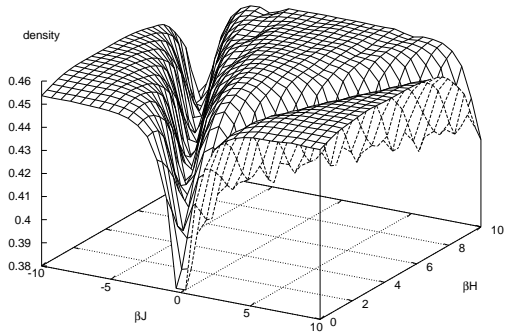}
\end{center}
\caption{\label{figd3d} Dependence of the MBD pile density on $\beta H$ and
$\beta J$. Observe trenches at borders between plateau  regions.}
\end{figure}

On fig.\ref{figd3d} one illustrates  in a 3D way the influence of the
interaction part of the hamiltonian, i.e. $\beta J$, on the density.
This sort of diagram allows us to emphasize that the  density is almost
the same everywhere, as mentioned above, but there are several minima:
the main density  variations occur along lines bordering  plateau
regions, lines which correspond to the equal probability lines mentioned
in the previous section. One can observe a set of trenches near these
borders between different growth (or packing) regions. In these
trenches, the density is markedly lower than in the immediate
neighborhood. Indeed such trenches correspond to the highest possible
rate of  sticking, thus to a condition for loose packing (see also
below).

\begin{figure} 
\begin{center}
\includegraphics[height=8cm, angle=-90]{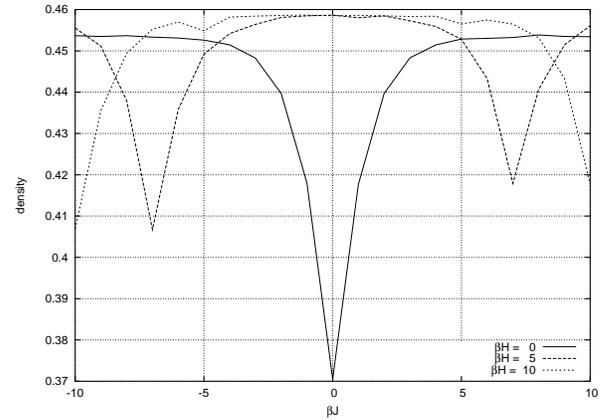}
\end{center}
\caption{\label{figdh} Dependence of the density on $\beta J$ for different
$\beta H$.} 
\end{figure}

 From Fig.\ref{figdh}, the value of the density is seen to be slightly
different for $\beta J > 0$ and for $\beta J < 0$,  but remains
qualitatively the same. These differences indicate that a little higher
density is obtained for ferromagnetic systems $\beta J > 0$ than for
antiferromagnetic ones, in particular when the external field is
different from $0$. Something similar had been found in studies on the
MDLA\cite{MDLA}. Further discussion on  this point is postponed for
after examination of the clusters in Sect.4.

\begin{figure}
\begin{center}
\includegraphics[height=8cm, angle=-90]{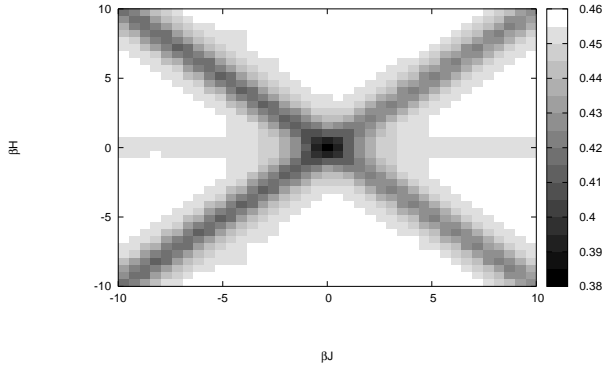}
\end{center}
\caption{\label{figd3dmap} Dependence of the density on $\beta H$ and
$\beta J$. Different gray colors correspond to different density levels with
''color'' scale indicated on the right.}
\end{figure}

Another illustration of the density dependence is exhibited on
Fig.\ref{figd3dmap} as a projection on the ($\beta J$, $\beta H$)
plane. The main trenches are observed, i.e. only $10$, out of all $32$
possible ones: they are located at  $\beta H=0, \quad \beta H=\pm \beta
J, \quad \beta H=\pm  2 \beta J$. One can  distinguish a trench for
positive and negative values of $\beta J$ and $\beta H$. The last one is
not clearly visible on Fig.\ref{figd3dmap}, but on Fig.\ref{figdh} one
can observe a small hollow near the trench $\beta  H = -\beta J$ for
$\beta J < 0$. Hollows are positioned along the trench $\beta
= - 2 \beta J$. Such structures (hollows) are artefacts due to 
resolution of the
simulation.

The $\beta H = \pm 2 \beta J$  trench  is very clearly seen on
Fig.\ref{figdh}; for $\beta H =5$  a small minimum is observed  for
$\beta J = \pm 5$, a symptom of the existence of a trench.

\begin{figure*}
\begin{center}
\includegraphics[width=9.5cm, angle=-90]{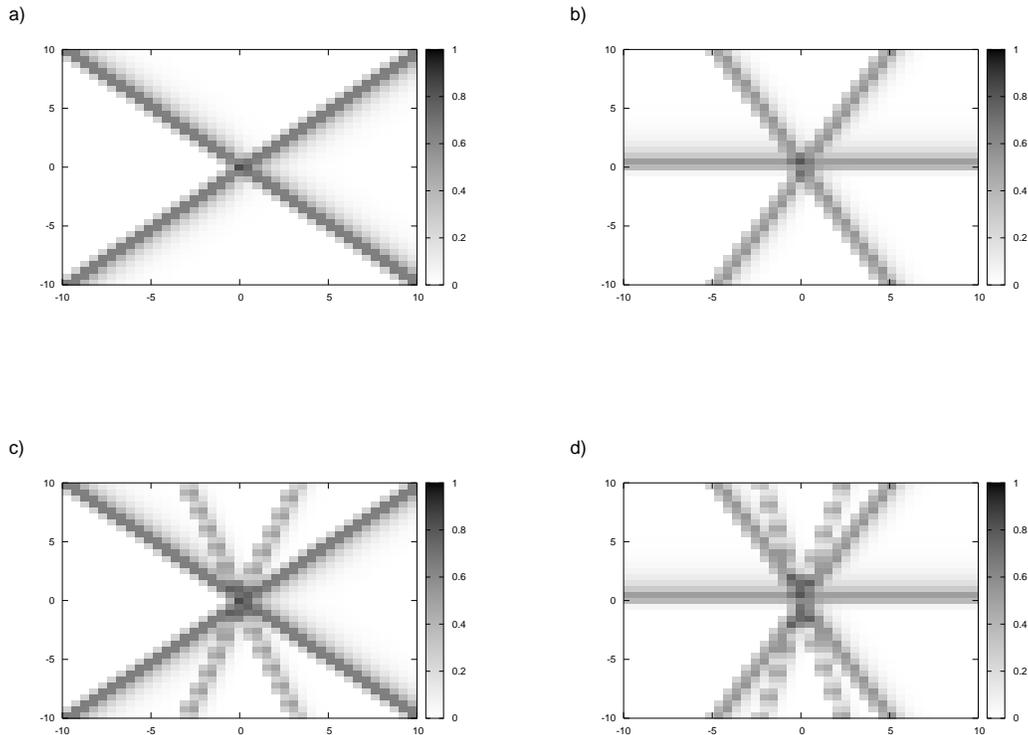}
\end{center}
\caption{\label{figprobstick} Theoretical maximum probability of
sticking a spin to a cluster for a cluster configuration with: (a) $1$
spin, (b) $2$ spins, (c) $3$ spins, (d) $4$ spins.}
\end{figure*}

To explain the behavior of the density due to $\beta J$ and $\beta H$
parameters let us compute the maximum probability of cluster sticking
for the  configuration having $n$ spins in a site neighborhood. The
results for each  configuration, with respect to the number of spins in
the perimeter are presented on Fig.\ref{figprobstick}. The discrete
color changes is real but the zig zag blurring is an artefact due to the
numerical sampling with $\beta J$ and $\beta H$.  Next, from this figure
it can be concluded that the biggest drop  in density occurs for $\beta
H= \pm \beta J$ indeed (Fig.\ref{figprobstick}a). The reason stems from
the  configurations which correspond to a case with a  small amount of
spins i.e. with $1$ or $3$ spins, thus to a possibly high relative
change in Ising energy. This observation further explains the existing
density trenches observed on Fig.\ref{figd3d} (or Fig.\ref{figd3dmap})
for $\beta H= \pm \beta J$.

A very similar conclusion pertaining to  the $\beta H = 0$ and $\beta H =\pm 2
\beta J$ lines can be drawn for the appropriate trenches in
Fig.\ref{figd3d} (or Fig.\ref{figd3dmap});  the drop is thereby smaller
because these configurations contain $2$ and $4$ spins, - starting from
a $4$ spin configuration does not obviously reduce the density after one
extra spin sticks to the cluster.

At each step, the available ''volume of interest'' on  the perimeter
corresponds to $7$ sites. For a configuration with the largest possible
number of  ''useful'' spins for sticking in the perimeter we have a
final $5$ spin configuration (4 + the dropping spin), whence the local
density  is at most $\frac{5}{7} \approx 0.71$. However this occurs at
most 6 times (out of 23), see Table I  leading to a rough estimate of
2(6/23)(5/7)=0.37 for the density.

On the other hand, configurations with $3$ and $4$ spins on the
perimeter, lead to an  increase in the density after spin sticking
roughly equal to $0.46$ as observed.
\begin{figure}
\begin{center}
\includegraphics[height=8cm, angle=-90]{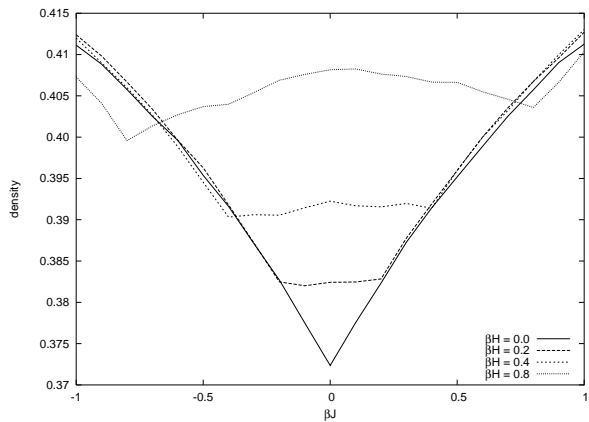}
\end{center}
\caption{\label{figkolodh} Dependence of the density on $\beta J$ for different
$\beta H$.}
\end{figure}

The behavior of the density in the vicinity of the zero field and zero
interaction case is of interest with respect to observe deviations from
the classical BD model. On Fig.\ref{figkolodh} it is seen that when the
field is different from $0$ the density is almost constant over some
$\beta J$ interval, the width of which depending on the field value,
i.e. the higher the field, the wider the interval.  Observe that the
density  value in the interval does not seem to be varying linearly with
the field. Notice that the  density dependence observed on
Fig.\ref{figdh} and its characteristic structures, i.e.  the trenches
are again observed for moderately high fields ($ca.$ 0.8).

\begin{figure}
\begin{center}
\includegraphics[height=8cm, angle=-90]{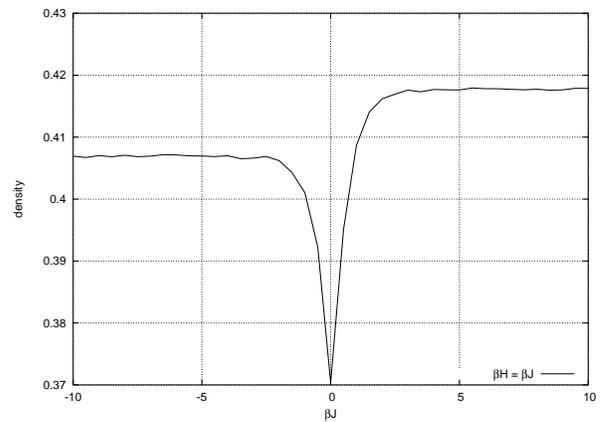}
\end{center}
\caption{\label{figprzek} Dependency of the density on $\beta J$ for $\beta
H=\beta J$ i.e., the deepest trench.}
\end{figure}

Finally, Fig.\ref{figprzek} exhibits the behavior of the density for $\beta H =
\beta J$ i.e., along the deepest trench (Fig.\ref{figd3d} or
Fig.\ref{figd3dmap}). It should be emphasized that the  $\beta J<0$
trench level has a lower density than  the  $\beta J>0$ one due to the
imbalance in sticking probabilities for preferred ferromagnetic or
antiferromagnetic-like configurations. Worth mentioning is that the
second deepest trench,  i.e. $\beta H = -\beta H$ has exactly the same
dependence like  that shown on Fig.\ref{figprzek}.

\subsection{ Magnetization}

In this subsection we  present results concerning the magnetization of
the cluster of spins. The magnetization is defined as:

\begin{equation}
M  = \frac{n_+ - n_-}{n_+ + n_-},
\end{equation}

where $n_+$ and $n_-$ are the  number of up and down spins respectively.
This quantity can be considered as a measure of the difference in grain
orientations in the packing.

\begin{figure}
\begin{center}
\includegraphics[height=8cm, angle=-90]{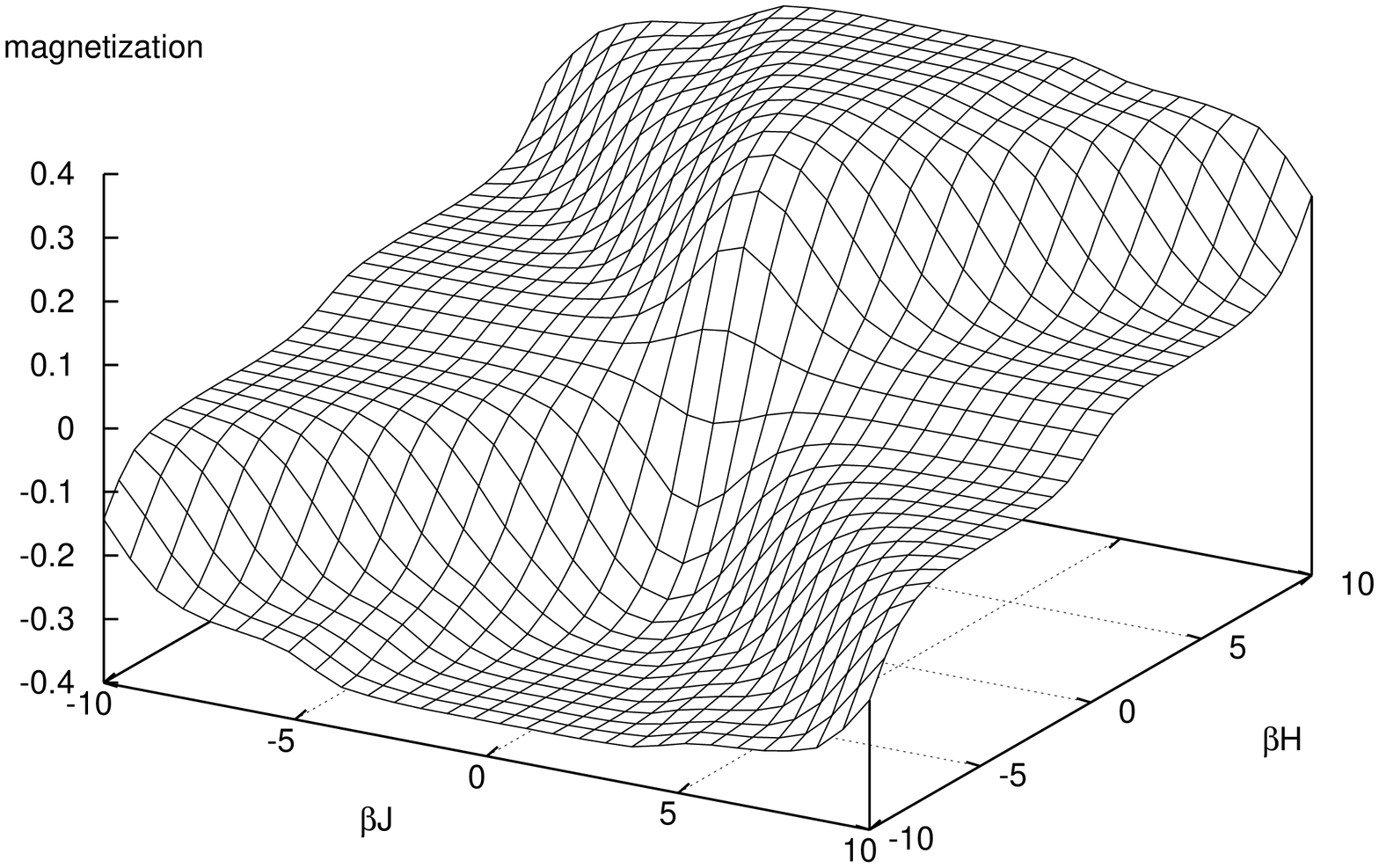}
\end{center}
\caption{\label{figm3d} Dependence of the magnetization on 
$\beta J$ and $\beta H$.}
\end{figure}

\begin{figure}
\begin{center}
\includegraphics[height=8cm, angle=-90]{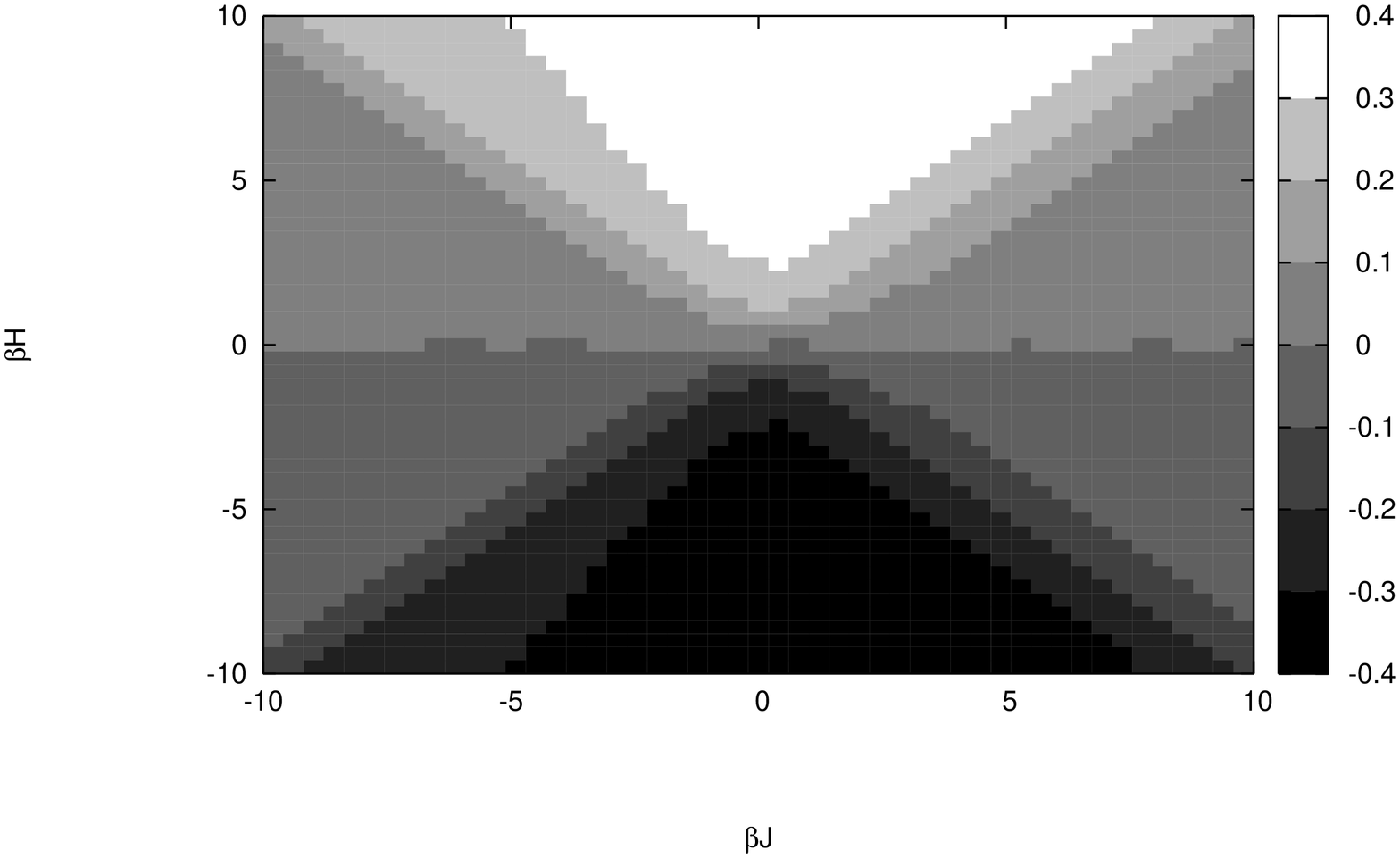}
\end{center}
\caption{\label{figm3dmap} Dependence of the magnetization on $\beta J$
for different $\beta H$, with a gray color scale as indicated.}
\end{figure}

As seen on Fig.\ref{figm3d} the magnetization dependence on $\beta H$
and $\beta J$ presents a sort of terraces and is slightly undulating, at
the borders of specific regions previously emphasized in the density
dependence discussion. Interestingly all expected dependences are better
visible on the magnetization than on the density pictures. In
particular, see Fig.\ref{figm3dmap}.

\begin{figure}
\begin{center}
\includegraphics[height=8cm, angle=-90]{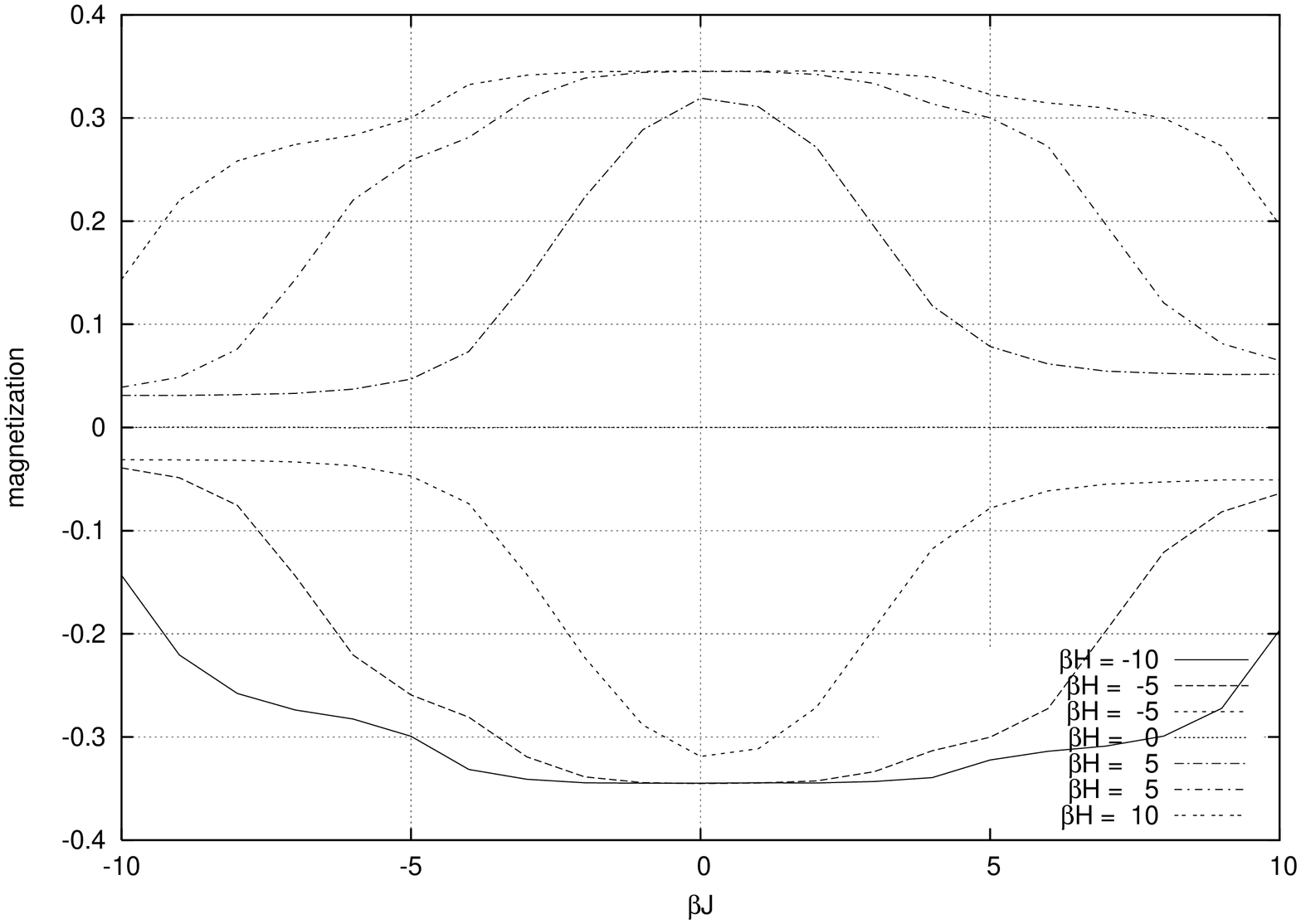}
\end{center}
\caption{\label{figmh} Dependence of the magnetization on $\beta J$
for different $\beta H$.}
\end{figure}

\begin{figure}
\begin{center}
\includegraphics[height=8cm, angle=-90]{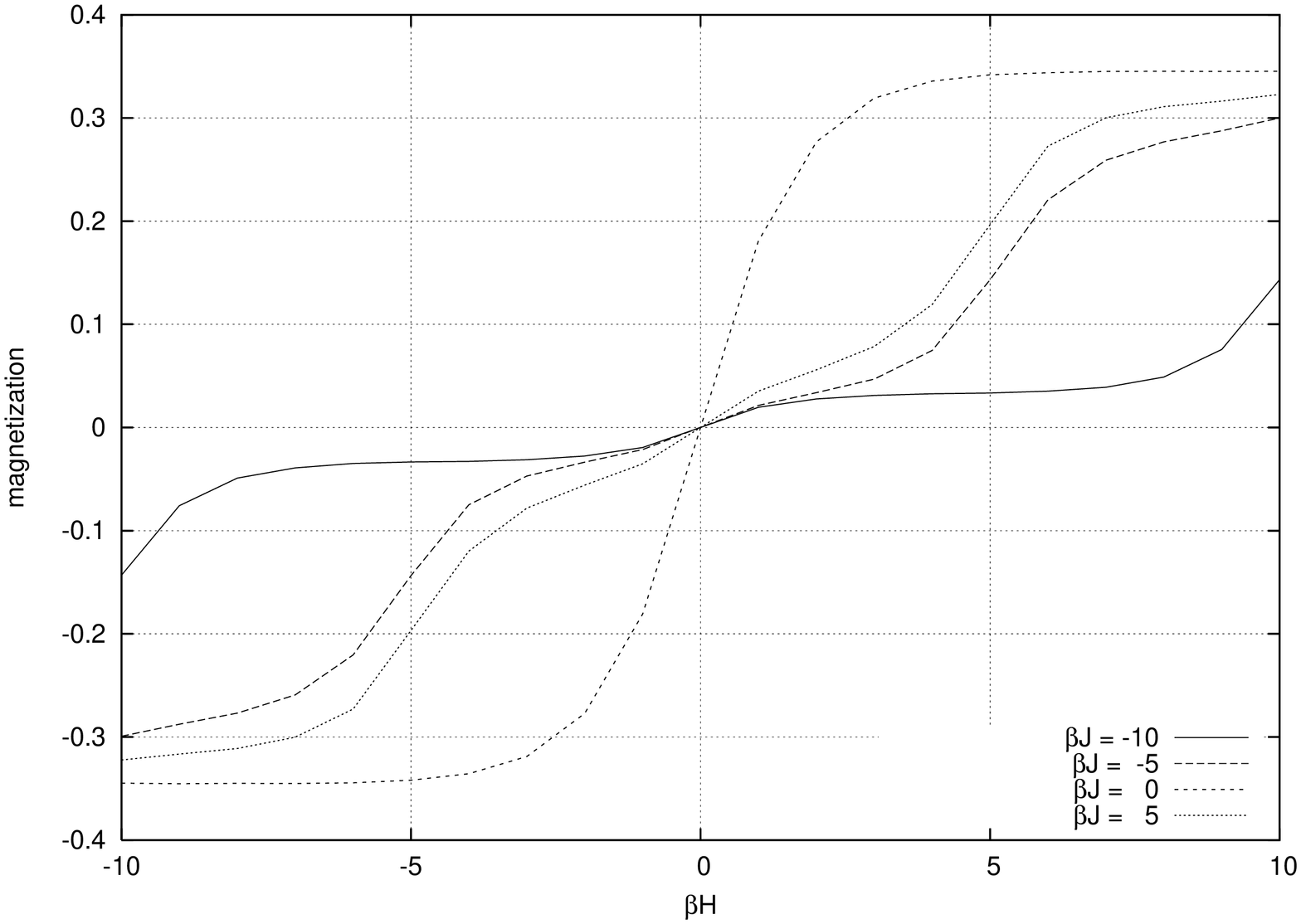}
\end{center}
\caption{\label{figmj} Dependence of the magnetization on $\beta H$
for different $\beta J$ values.}
\end{figure}

Fig.\ref{figmh} and Fig.\ref{figmj}  present  the behavior of the
magnetization for different $\beta J$ and $\beta H$ values. The maximum
magnetization occurs for $\beta J=0$  and no field. For a finite field
the maximum is rather broad. Terrace  structures are seen like in
Fig.\ref{figm3d} for the  density. There are about $6$ terraces with
different values of the magnetization. Each terrace occurs in ranges
like those of the regions  previously mentioned. It is emphasized that
the magnetization terrace levels {\em differs} in the antiferromagnetic  and
the  ferromagnetic coupling regions, i.e.  for $\beta J<0$ and $\beta J
>0$, - the level height depending on the field sign.

This is well stressed through  figure (Fig.\ref{figmj}) which  exhibits
the dependence of the magnetization on $\beta H$ for different $\beta
J$.  Notice that for the case  {\it without spin-spin interaction},  one
obtains a kind of saturation  at a value $ca$. $\pm 0.36$. Observe the
surprising form of the ($M,H$) curve, as for classical soft magnets. For
a finite field, the magnetization does not saturate like in the case of
zero field, for the  values that we have investigated, but $M$ saturates
creating stair-like structures.

\subsection{ Magnetic susceptibility}

The magnetic susceptibility of the clusters can be obtained by numerically
differentiating  the magnetization $M$ over the field $H$, at fixed $J$ or $H$,
i.e.

\begin{center}
\begin{equation}
\chi_J = \left. \frac{d M}{d H}\right|_J, \quad
\chi_H = \left. \frac{d M}{d J}\right|_H.
\end{equation}
\end{center}

\begin{figure}
\begin{center}
\includegraphics[height=8cm, angle=-90]{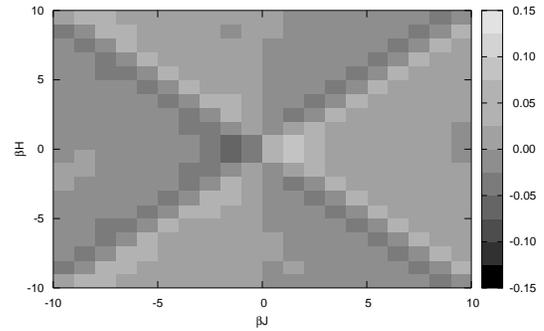}
\end{center}
\caption{\label{figmagsus} Dependence of the magnetic susceptibility
on $\beta J$ and $\beta H$.}
\end{figure}

The dependence is presented on Fig.\ref{figmagsus}. The highest
susceptibility occurs for regions where $\beta H=\beta J=0$ and the main
trenches.  Other regions have a rather relatively small susceptibility.
The results can be understood through the role of the interaction
between spins which as usual  induces a  drop in the magnetic
susceptibility of the materials. On the other hand,  the variation of
the orientation difference ($M$) of the grains in a packing with
respect to an external (wind or) field should be an interesting
experimental test.

\subsection{ Compressibility}

\begin{figure}
\begin{center}
\includegraphics[height=8cm,angle=-90]{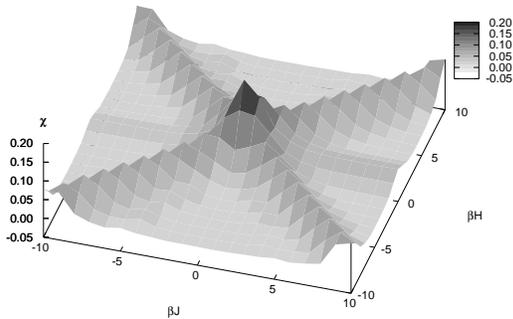}
\end{center}
\caption{\label{figcomp} Dependence of the compressibility
on $\beta J$ and $\beta H$.}
\end{figure}

Similarly the compressibility (Fig.\ref{figcomp}) at fixed $H$ or $J$, is

\begin{equation}
\kappa_H = \left. -\frac{1}{G} \frac{d G}{d J}\right|_H, \quad
\kappa_J = \left. -\frac{1}{G} \frac{d G}{d H}\right|_J,
\end{equation}

where $G$ is the density. The displayed  data is rather blurred  because
of the limited amount of data for numerical differentiation  near the
trenches and in the region of the standard BD, in particular.
Nevertheless this figure indicates some mild  variation due to internal
competition and external conditions.  Let us also recall  that while the
susceptibility is singular in spin glass models of
compaction\cite{Coniglio} at the critical percolation value, the
compressibility seems to remain finite.\cite{Coniglio,JPI1996} Our model
indicates the same, as experimentally or numerically  observed. Further
experimental and numerical considerations should be given to this point.

\section{ PILE STRUCTURES}

In this section we present some examples of typical clusters created by the MBD
in specific regions, as observed and discussed here above. The size of the
lacunes in each cluster allows some emphasis and contrasting.

\subsection{ Typical clusters}

\begin{figure*}
\begin{center}
\includegraphics[height=13cm]{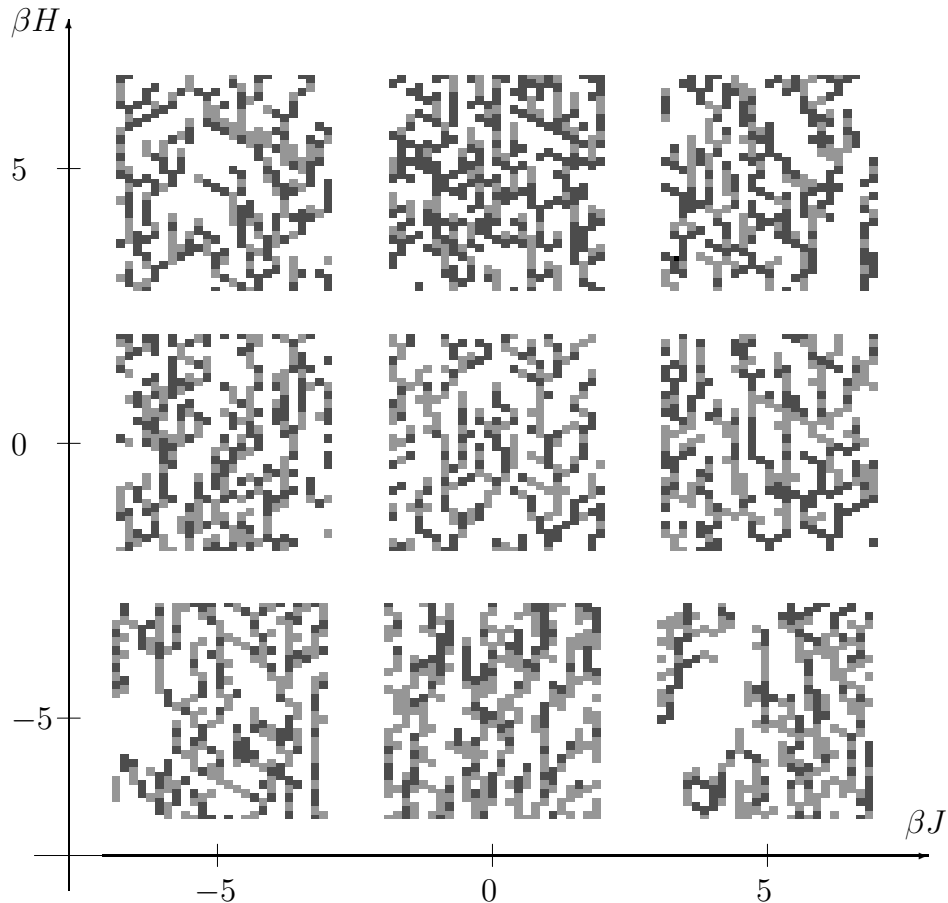}
\end{center}
\caption{\label{figexamples}
Typical clusters. The ''colors'' correspond to the spin's sign.}
\end{figure*}

In Fig.\ref{figexamples}, 9 clusters from 9 different growth regions  are
shown . The central cluster has the smallest density: it corresponds to
the case when there is no interaction between spins and no field, i.e.
it is the standard BD model\cite{BDM}.

On the other hand the clusters for $\beta J=0$ and $\beta H = \pm 5$ have the
highest density;  they illustrate regions where the density is
saturating (above
$0.44$, see Fig.\ref{figdj}). It should be mentioned that the results are
symmetrical with respect to the field sign.

For a finite $J$ interaction between spins, a quite different behavior
of the clusters is observed. The clusters corresponding to the trenches
($\beta J= \pm 5$ and $\beta H= \pm 5$) have a smaller density
(ca.$0.41$) than those outside the trenches (see for example clusters
with $\beta J= 0$ and $\beta H= \pm 5$). Also observe, in
Fig.\ref{figexamples}, field-free grown clusters and  differences between
$\beta J < 0$ and $\beta J > 0$ clusters. When $\beta J >  0$, the spins
show a tendency toward similar sign spin ''domains''. In the
antiferromagnetic-like region, i.e. where $\beta J <0$, adjacent  spins
have more often opposite directions, -- a  cluster with $\beta J<0$ and
$\beta  H = 0$ is a typical example. This allows us to emphasize that
the internal  competition leads to different cooperative phenomena in
cluster packing.\cite{Faraudo}

\subsection{ Fractal dimension}

\begin{figure}
\begin{center}
\includegraphics[height=8cm, angle=-90]{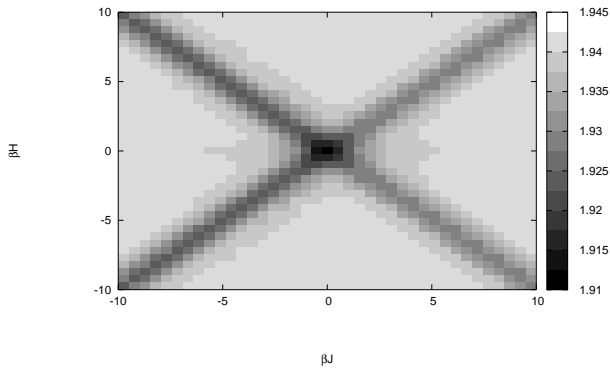}
\end{center}
\caption{\label{figdimwhole} The fractal dimension for the whole cluster, i.e.
not recognizing the spin signs.}
\end{figure}

\begin{figure*}
\begin{center}
\includegraphics[height=6cm,angle=-90]{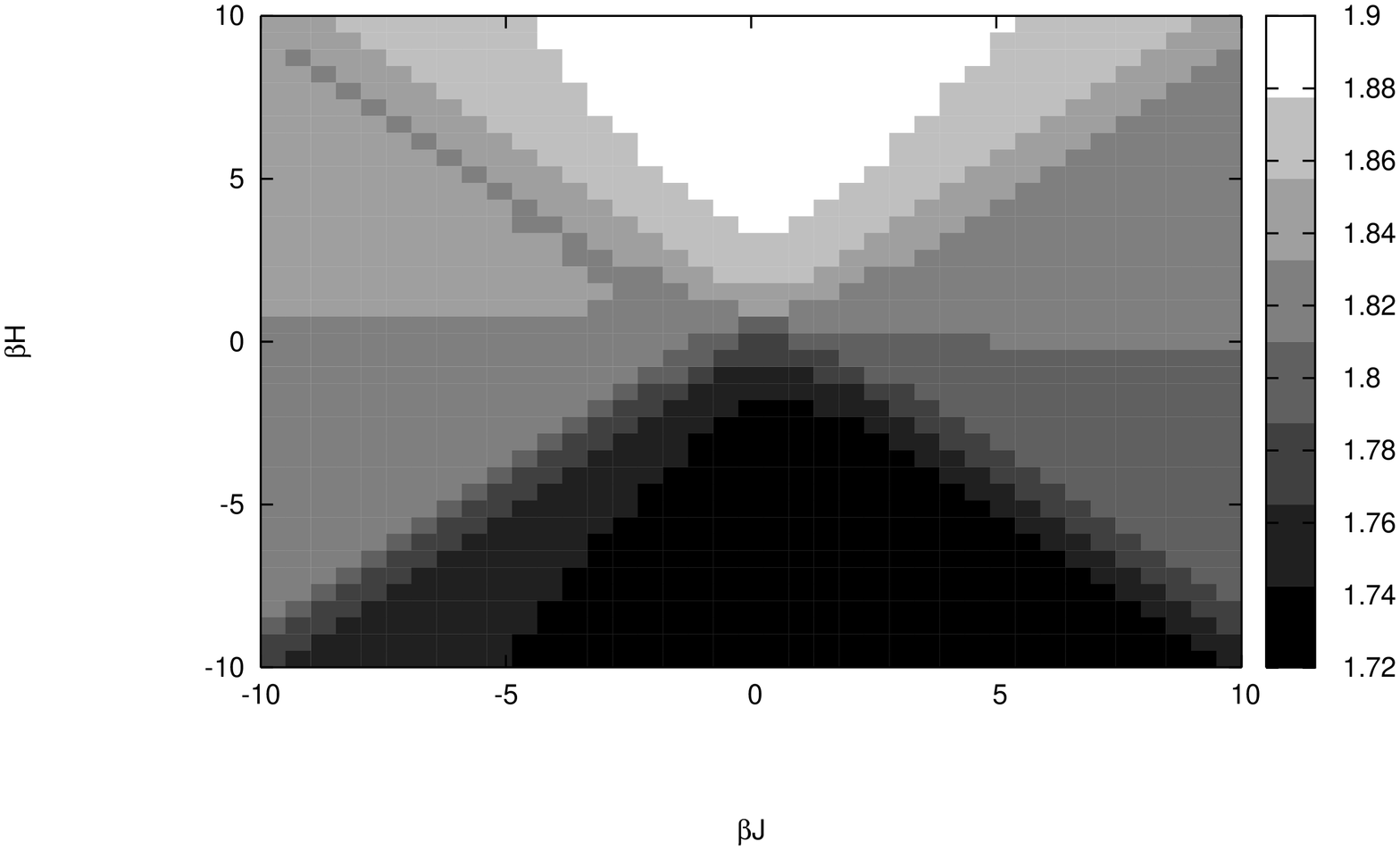}
\includegraphics[height=6cm, angle=-90]{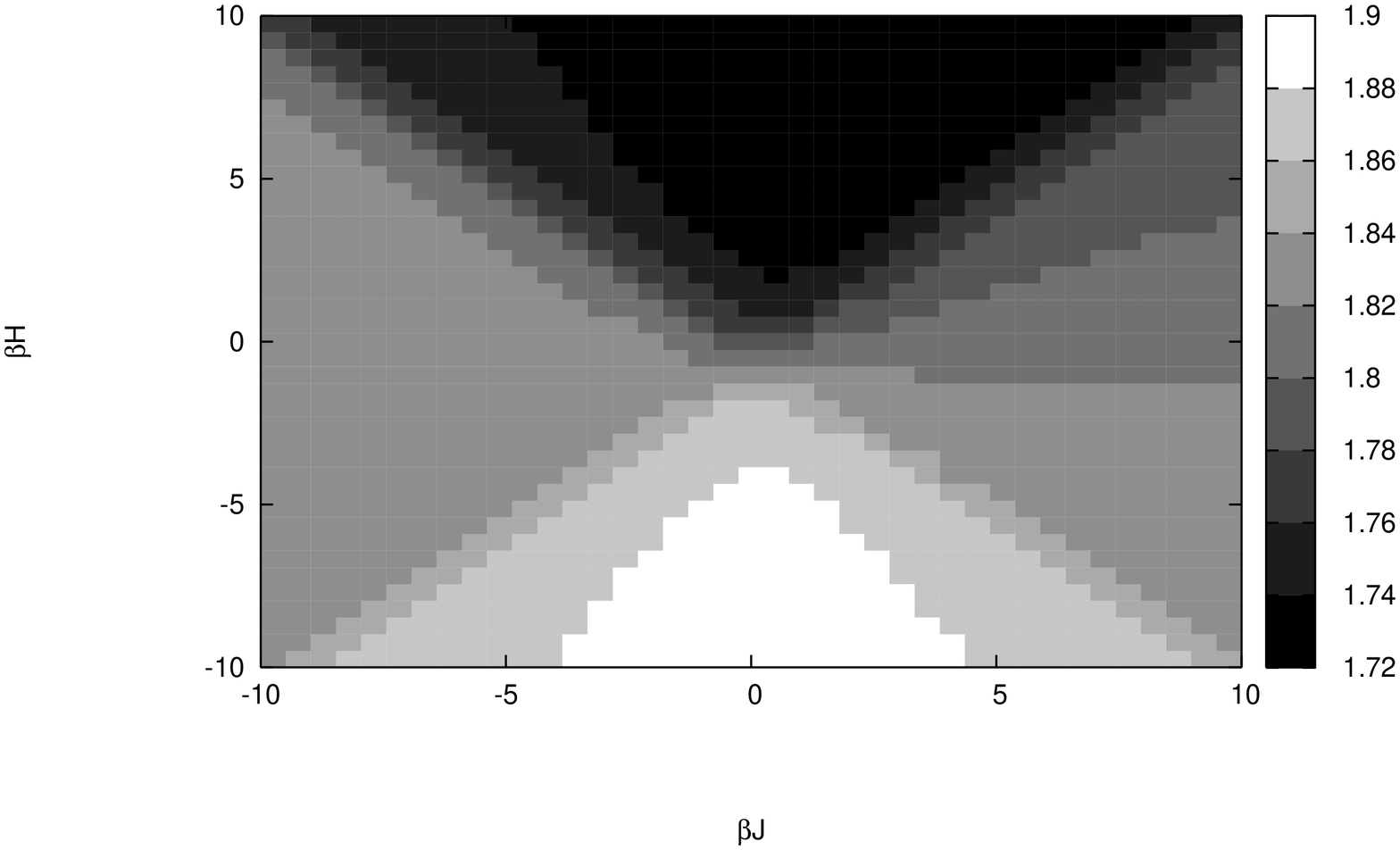}
\end{center}
\caption{\label{figdimupdown} The
fractal dimension for the granular cluster when we recognize the spin
signs.}
\end{figure*}

For further relating the model and our investigations to granular
piles, it is of interest to check  the fractal
dimension\cite{refA1,WestDeering} of  the piles in the different
parameter regions. The (box counting) technique\cite{refA1,WestDeering}
consists in covering, without  overlapping, the whole cluster by squares
of the same size, and computing the number of squares which have at
least one spin up (or down), for different square sizes. We have
distinguished between the fractal dimension of the cluster of up and
down spins. When computing the fractal dimension of the whole cluster it
was simply checked whether there was at least one spin and its sign in
the relevant  square. From the best  linear fit to the data, i.e.
$-\log(\mbox{square size})$ vs. $\log(\mbox{number of squares with a
spin})$, the fractal dimension is obtained through the slope. The
results are reported in Figs.\ref{figdimwhole}-\ref{figdimupdown}.

Let us discuss the results corresponding to different regions, either
not distinguishing over the spin sign (Fig.\ref{figdimwhole}), or on the
contrary considering down or up cases (Fig.\ref{figdimupdown}a or
Fig.\ref{figdimupdown}b,respectively). The fractal dimension in every
case ranges from 1.91 to 1.95, therefore is about equal to $2$, taking
into  consideration the error bars. This value is similar to what is
found in classical BDM\cite{Herrmann,BDM} and in the rain
model\cite{Cerdeira}. Surprisingly the lowest fractal dimensions are found in
the above for  the trenches and in  particular for the classical BDM.

\section{ CONCLUSION}

We have presented  a nonequilibrium ballistic deposition model  with one
degree of freedom per entity, degree which can be coupled to an external
field. We have examined the cluster properties emphasizing the existence
of two order parameters,  since two characteristic field ($J$ and $H$)
are intrinsic to the model. This model can serve to describe in a first
approximation the deposition of a distribution of grains, distribution
characterized by one intrinsic parameter which can be coupled to an
external field.  The degree of freedom can be either the anisotropy
factor or the surface roughness or an electrostatic imbalance of a
grain, - the corresponding field being immediately thought of.

For the sake of such an extension of usual deposition models, the degree
of freedom has been called ''spin''. We have simulated the
nonequilibrium deposition in a finite size 2D vertical box, admitting
that grains flow down along  linear trajectories on a triangular
lattice. The "quenching" of the degree  of freedom on the cluster leads
to branching or compactness and moreover to  combined geometric and
physical regions at specific ''field'' and ''spin-spin  interaction''
values. This was seen through the calculation of the ''density'' and the
so called ''magnetization''. Different cluster regimes were expected
according  to the (spin sticking) packing rule.

We have investigated a box geometry and a triangular underlying lattice.
The ranges in density and magnetization are limited, but features exist
resulting from competitive nonequilibrium growth/deposition processes.
Minima in density occur along specific sticking probability lines.
Slight differences  exist whether the ''spin-spin interaction energy''
is positive or negative. The fractal dimension of clusters whatever the
type of grains and the parameter sign or values is however rather
trivial and equal to 2. This differs  markedly from what was found in
the MDLA model starting from a central seed in which  both in the
ferromagnetic interaction regions, and the AF regions, the cluster
morphology was dendritic with an important thickening of the branches
and  the  fractal dimension ranging from $1.68$ to $1.99$. Instead of a
critical a value $\beta  J_c$ a set of values depending on the external
field divide the parameter plane in regions, with trenches and plateaus.
Thus, a spreading phenomenon is avoided  around such $\beta J_c(H)$. It
is of interest to further examine whether this has  interesting
consequences in granular deposition situations.\cite{Chowhan}

The connection between this model and granular matter systems suggests
some experimental work. In the MBD the spin can be interpreted as a
rotation or defining a direction process. The coupling constant can be
mapped to  a mechanical friction energy, the magnetic field to gravity
or wind pressure. It  is known that packing is impaired by static
electrical charges. Such effects may be  considered in the above
framework.

{\bf Acknowledgments} 

KT is supported through an Action de Recherche Concert\'ee Program of 
the University
of Li$\grave e$ge (ARC 02/07-293). Comments by R. Cloots, L. Delattre, A.
P\c{e}kalski, N. Vandewalle and other coworkers of the ARC program  on
granular materials at ULg are greatly appreciated.

\end{document}